\documentclass[english,a4paper,14pt]{article}
\usepackage{babel,amssymb,graphicx,hyperref}

\advance\textheight \topskip \textwidth 35.5pc \oddsidemargin 20pt
\begin{document}
\title {Cosmological status of Lagrangian theory\\ of density perturbations}

\author{V.~Strokov$^{1,2}$\footnote{{\bf e-mail}: strokov@asc.rssi.ru},
\\
$^1$ \small{\em Astro Space Center of the P.~N.~Lebedev Physical Institute of RAS} \\
\small{\em 117997 Moscow, ul. Profsoyuznaya, 84/32} \\
$^2$ \small{\em Moscow Institute of Physics and Technology,
Department of General and Applied Physics} \\
\small{\em 141701 Dolgoprudny, Institutskiy per., 9} }

\date{}
\maketitle{}
\begin{abstract}
We show that hydrodynamical and field approaches in theory of
cosmological scalar perturbations are equivalent for a single
medium. We also give relations between notations introduced by
V.~Lukash, J.~Bardeen, J.~Bardeen~et~al. and G.~Chibisov and
V.~Mukhanov.
\end{abstract}

\maketitle

 \clearpage

\newpage

\section{Introduction}
In the linear theory of primordial cosmological perturbations the
metric tensor $g_{\mu \nu}$ and energy-momentum tensor of matter
$T_{\mu\nu}$ are splitted into a background part and a
perturbation part: $g_{\mu \nu}=g^{(0)}_{\mu \nu}+h_{\mu \nu}$,
$T_{\mu\nu}=T_{\mu\nu}^{(0)}+\delta T_{\mu\nu}$. Evidently, this
splitting into background and perturbation is ambiguous. Making
small coordinate transformations we obtain a different background
and a different perturbation. And in a given reference frame we
can get "perturbations" which are actually not physical and are
just a result of the chosen reference frame. In order to avoid
such phantom perturbations, formalism of gauge-invariant variables
is used. Equations which describe the evolution of a
gauge-invariant variable come from the perturbed part of Einstein
equations and properties of matter. Clearly, Einstein equations
solely are insufficient to solve the dynamical problem as they
just relate perturbations of the metric tensor with those of the
energy-momentum tensor without specifying any physics. In order to
obtain a dynamical equation we need a physical relation. For
example, it can be a relation between matter density and pressure
perturbations which corresponds to the \textit{hydrodynamical}
approach. Also we can start from a general matter Lagrangian of
$\varphi$-field (which serves as a 4-velocity potential of
matter), and this corresponds to the \textit{field} approach. The
former is usually associated with equation of state of matter,
while the latter is with models of inflation. In either case it
allows one to get dynamics and a Lagrangian of perturbations which
appear to be independent on the initial assumptions.

Further we set out the main results of the Lagrangian theory of
scalar perturbations~\cite{lukash} without derivation. We show the
equivalence of hydrodynamical and field approaches for a single
medium. Then we show how notations of Lukash~\cite{lukash},
Bardeen~\cite{Bardeen 1}, Bardeen et al.~\cite{bardeen} and
Chibisov and Mukhanov~\cite{mukhanov} are related.

\section{Gauge-invariant formalism in a nutshell}
Below we work with the background Friedmann-Robertson-Walker
geometry:
\begin{equation}
\label{geometry}
\begin{array}{c}
ds^{2}=dt^{2}-a^{2}dx_{i}dx^{i}=a^{2}(d\eta^{2}-dx_{i}dx^{i}),\\
T_{\mu\nu}=(\varepsilon+p)u_{\mu}u_{\nu}-pg_{\mu\nu}.
\end{array}
\end{equation}
The background metric tensor is
$g_{\mu\nu}=diag(1,-a^{2},-a^{2},-a^{2})$, matter density
$\varepsilon$ and pressure $p$ are functions of time, and the
background 4-velocity is~$u^{\mu}=(1,0,0,0)$, the speed of
light~$c=1$, and $\eta$ is conformal time,~${d\eta=dt/a}$.
Hereafter we omit the superscript~$(0)$ for background quantities.

The basic equations for the scale factor~$a(t)$ are the Friedmann
equations:
\begin{equation}
\label{fridmann2} H^{2}=\frac{8\pi G}{3}\varepsilon,
\end{equation}
\begin{equation}
\label{fridmann1}
\gamma\equiv-\frac{\dot{H}}{H^{2}}=\frac{3}{2}\left(1+\frac{p}{\varepsilon}\right),
\end{equation}
where $H=\dot{a}/a=a'/a^{2}$ is the Hubble parameter, and~$G$ is
the gravitational constant. The dot and the prime stand for the
derivative with respect to physical time~$t$ and conformal
time~$\eta$, respectively.

Generally, scalar type metric perturbations are constructed using
four potentials~\cite{fieldtheory},~$A$,~$B$,~$C$ and $D$:
\begin{equation}
\label{metric-perturbations} h_{\mu \nu}=\left(
\begin{array}{cc}
2D & C_{,i} \\
C_{,i} & 2a^{2}(A\delta_{ij}+B_{,ij})
\end{array}\right).
\end{equation}
The potential~$A$ is actually a perturbation of the scale
factor:~$A=-\delta a/a$.

Perturbation of the energy-momentum tensor is presented via other
four potentials, $\upsilon$, $\delta\varepsilon$, $\delta p$ and
$E$:
\begin{equation}
\begin{array}{c}
\delta T^{0}_{0}=\delta\varepsilon, \\
~ \\
\delta T^{0}_{i}=(\varepsilon+p)\upsilon_{,i}, \\
~ \\
-\delta T^{i}_{j}=\delta p\delta_{ij}+(\varepsilon+p)\sigma_{ij},
\end{array}
\end{equation}
$$
\displaystyle\sigma_{ij}=\frac{1}{2a^{2}H^{2}}(E_{,ij}-\triangle
E\delta_{ij}),~~~\sigma_{i,j}^{j}=0,
$$
where $E$ presents anisotropic stresses, and~$\upsilon$ is the
3-velocity potential:
\begin{equation}
\label{four-vel} u_{\mu}=(1+D, \upsilon_{,i}).
\end{equation}

Thus, we have four gravitational potentials~$A$,~$B$,~$C$,~$D$ and
four matter potentials~$\upsilon$,~$\delta\varepsilon$,~$\delta
p$,~$E$. All of them but~$E$ are not gauge-invariant. By small
coordinate transformations $x^{\mu}\rightarrow x^{\mu}+\xi^{\mu}$
the potentials get changed. Two of these eight potentials are
arbitrary, they correspond to a gauge choice (an arbitrary vector
in scalar representation~$\xi_{\mu}=Fu_{\mu}+H_{,\mu}$). It is
possible to construct some gauge-invariant combinations of the
potentials. All such combinations constitute an infinite set.

The
potentials~$A$,~$B$,~$C$,~$D$,~$\upsilon$,~$\delta\varepsilon$,~$\delta
p$,~$E$ are not independent. They are linked through the
first-order expansion of the Einstein equations
\begin{equation}
\label{einstein} \delta G^{\mu}_{\nu}=8\pi G\delta T^{\mu}_{\nu}.
\end{equation}

The natural gauge-invariant combination is that of the
gravitational potential~$A$ and the velocity potential~$\upsilon$
which is called the $q$-scalar~\cite{lukash}:
\begin{equation}
\label{qsc} q=A+H\upsilon.
\end{equation}

The inverse transformations of the~$q$-field to the original potentials are as
follows:
\begin{equation}
\label{inverse-transformations1}
\upsilon=\displaystyle\frac{q-A}{H}, ~~~~ \delta p_{c}\equiv\delta
p-\dot{p}\upsilon=\frac{\varepsilon+p}{H}\dot{q},
\end{equation}
\begin{equation}
\label{inverse-transformations2} a^{2}\dot{B}-C=\frac{A-\Phi}{H},
~~~~ D=\gamma q-\frac{d}{dt}\left(\frac{A}{H}\right),
\end{equation}
\begin{equation}
\label{inverse-transformations3}
\delta\varepsilon_{c}\equiv\delta\varepsilon-\dot{\varepsilon}\upsilon=\displaystyle\frac{\triangle\Phi}{4\pi
Ga^{2}}, ~~~~\Phi=\frac{H}{a}\int{a\gamma(q+E) dt},
\end{equation}
where~$\delta p_{c}$~and~$\delta\varepsilon_{c}$ are
gauge-invariant variables of pressure and energy density
perturbation, respectively. The first equation
in~(\ref{inverse-transformations3}) is, in fact, the relativistic
Poisson equation. From the inverse transformations it can
explicitly be seen that~$\Phi$~and~$E$ are gauge-invariant
potentials.

The previous analysis is common and does not depend on any matter
physics. However, in order to introduce dynamics we need some
additional relation between matter quantities (e.g.
between~$\delta p_{c}$ and~$\delta\varepsilon_{c}$), that is, we
need to specify some physics. We have two possibilities: either we
can use the \textit{hydrodynamical} approach to relate~$\delta
p_{c}$ and $\delta\varepsilon_{c}$ or the \textit{field} approach,
i.e. to admit some form of the matter Lagrangian for 4-velocity
potential of the medium. Further it is shown that the both
approaches are equivalent in this problem. Further on we consider
a single medium. Also, we suppose absence of anisotropic stresses,
therefore,
\begin{equation}
\begin{array}{cc}
E=0, & \Phi=\displaystyle\frac{H}{a}\int a\gamma qdt.
\end{array}
\end{equation}

\textbf{Hydrodynamical approach}. In the hydrodynamical approach
we assume
\begin{equation}
\label{delta-p-c} \delta p_{c}=\beta^{2}(t)\delta\varepsilon_{c},
\end{equation}
where~$\beta(t)$ is a function of time. Hence
from~(\ref{inverse-transformations1})~and~(\ref{fridmann2}),
\begin{equation}
\label{delta-e-c}
\delta\varepsilon_{c}=\alpha^{2}H\dot{q},~~~\alpha^{2}=\frac{\gamma}{4\pi
G\beta^{2}}.
\end{equation}
Relation (\ref{delta-p-c}) means that there is only one medium and
we describe its perturbations. As soon as~(\ref{delta-p-c}) is
valid equations~(\ref{inverse-transformations3})
and~(\ref{delta-e-c}) immediately give:
\begin{equation}
\gamma\beta^{-2}a^{3}\dot{q}=\int{a\gamma\triangle qdt}.
\end{equation}
After differentiation the last equation gives equation describing
the evolution of {$q$-scalar}:
\begin{equation}
\label{qscalar} \ddot{q}+\left(3H+2\frac{\dot{\alpha}}{\alpha}
\right) \dot{q}-{\left(\frac{\beta}{a}\right)}^{2}\triangle q=0.
\end{equation}
The equation~(\ref{qscalar}) corresponds to the
action~\cite{lukash}
\begin{equation}
\label{lagrangian-q}
\begin{array}{c}
S[q]=\displaystyle\frac{1}{2}\int{\alpha^{2}\left(\dot{q}^{2}-\left(\displaystyle\frac{\beta}{a}\right)^{2}q_{,i}q^{,i}\right)a^{3}dtd^{3}x}=\\
{}\\
=\displaystyle\frac{1}{2}\int{(\alpha
a)^{2}\left({q'}^{2}-\beta^{2}q_{,i}q^{,i}\right)d\eta d^{3}x}.
\end{array}
\end{equation}

Since the backward path from equation to a Lagrangian defines the
Lagrangian to a factor before it, we can see that
(\ref{lagrangian-q}) has the right coefficient if we look at it in
some asymptotic limit, e.g. in the limit of small scales (the
sound wave frequency~$\omega\gg H$ and sound
velocity~$c_{s}\simeq\beta$). In this approximation~$q\simeq
H\upsilon$,~$\dot{q}\simeq H\dot{\upsilon}$ and
$\delta\varepsilon_{c}\simeq\delta\varepsilon$. Using the
relations:
\begin{equation}
\begin{array}{cc}
\displaystyle\frac{\delta\varepsilon}{\varepsilon+p}=\displaystyle\frac{\dot{\upsilon}}{c_{s}^{2}},&
\displaystyle\frac{\nabla\upsilon}{a}=-{\mathbf{v}},
\end{array}
\end{equation}
where~${\mathbf v}$ is hydrodynamical velocity in a sound wave, we
have the following chain of equalities:
\begin{equation}
\begin{array}{c}
L[q]=\displaystyle\frac{1}{2}(a\alpha)^{2}\left({q'}^{2}-\beta^{2}q_{,i}q^{,i}\right)=\displaystyle\frac{1}{2}(\alpha
aH)^{2}\left({\upsilon'}^{2}-c_{s}^{2}(\nabla\upsilon)^{2}\right)=\\
{}\\
=\displaystyle\frac{a^{4}}{2}\left(c_{s}^{2}\displaystyle\frac{\delta\varepsilon^{2}}{\varepsilon
+p}-(\varepsilon +p){\mathbf{v}}^{2}\right).
\end{array}
\end{equation}
The corresponding comoving volume energy density is
\begin{equation}
{\mathcal{E}}=\displaystyle\frac{1}{2}\left(c_{s}^{2}\displaystyle\frac{\delta\varepsilon^{2}}{\varepsilon
+p}+(\varepsilon +p){\mathbf{v}}^{2}\right).
\end{equation}
The last expression is exactly the energy density in a sound
wave~\cite{hydrodynamics},~\cite{lukash}.

\textbf{Field approach}. The Universe filled with a scalar
field~$\varphi$. The relation to the 4-velocity~(\ref{four-vel})
is
\begin{equation}
u_{\mu}=\frac{\varphi_{,\mu}}{w}, \label{4-velocity}
\end{equation}
where~${w^{2}=\varphi_{,\mu}\varphi_{,\nu}g^{\mu\nu}}$. The
Lagrangian density of the scalar field can be taken in a quite
arbitrary form~\cite{lukash},~\cite{lukash-lectures}:
\begin{equation}
\label{lagrangian-phi} {\mathcal L}={\mathcal L}(\varphi, w).
\end{equation}
From the matter Lagrangian~(\ref{lagrangian-phi}) we obtain the
energy-momentum tensor:
\begin{equation}
\varepsilon=nw-\mathcal{L},~~~~p=\mathcal{L},~~~~n\equiv\displaystyle\frac{\partial{\mathcal
L}}{\partial w},~~~~\upsilon=\frac{\delta\varphi}{\dot{\varphi}}
\end{equation}
and hence, the following relation between~$\delta\varepsilon_{c}$
and $\delta p_{c}$:
\begin{equation}
~~~~\frac{\delta\varepsilon_{c}}{\delta p_{c}}=\frac{n_{,w}w\delta
w-n_{,w}\dot{w}\delta\varphi}{n\delta
w-n\dot{w}w^{-1}\delta\varphi}=\frac{n_{,w}w}{n}=\displaystyle\frac{w}{n}\displaystyle\frac{\partial^{2}{\mathcal
L}}{\partial w^{2}}\equiv c_{s}^{-2}(\varphi, w).
\end{equation}

For the linear perturbations the function $c_{s}^{-2}(\varphi, w)$
is taken in the zero order and turns into a function of time
$c_{s}^{-2}=c_{s}^{-2}(t)$. This proves that the background
functions are identical,~$c_{s}=\beta$, and, thus, both ways,
(\ref{delta-p-c}) and (\ref{lagrangian-phi}), of deriving
equation~(\ref{qscalar}) are identical as well.

In the field approach one can obtain the Lagrangian describing
perturbations by expanding straightforwardly  the action for
gravitating scalar field to the second order in perturbation. The
action is standard:
\begin{equation}
\label{action} S[\varphi, g_{\mu\nu}]=\int{({\mathcal L}-\frac{1}{16\pi
G}R)(-g)^{1/2}d^{4}x},
\end{equation}
where~$R$ is scalar curvature. Perturbing the variables to the
linear order, ~${g_{\mu\nu}\rightarrow
g_{\mu\nu}+h_{\mu\nu}}$,~${\varphi\rightarrow\varphi+w\upsilon}$,
and decomposing~(\ref{action}) up to the second order terms we
obtain the action for perturbations~(total divergency terms are
omitted):
\begin{equation}
\begin{array}{c}
\label{expansion}
\delta^{(2)}S[\upsilon,h_{\mu\nu}]=-\displaystyle\frac{1}{64\pi G}\int{(\overline{h}_{\sigma\beta;\alpha}\overline{h}^{\sigma\beta;\alpha}-2\overline{h}^{\alpha\beta}{}_{;\sigma}\overline{h}^{\sigma}{}_{\alpha;\beta}-\frac{1}{2}\Box\overline{h})(-g)^{1/2}d^{4}x}-\\
{}\\
\qquad
{}-\displaystyle\frac{1}{4}\int{(\displaystyle\frac{1}{16\pi G}R-
{\mathcal
L})(\overline{h}^{\mu}_{\nu}\overline{h}^{\nu}_{\mu}-\displaystyle\frac{1}{2}\overline{h}^{2})(-g)^{1/2}d^{4}x}+\\
{}\\
\qquad
{}+\displaystyle\frac{1}{2}\int{nw\left[\nu\upsilon\overline{h}-2u_{\mu}\upsilon_{\nu}\overline{h}^{\mu\nu}+\chi^{2}(c_{s}^{-2}-1)+m^{2}\upsilon^{2}+2\Gamma\upsilon\chi\right](-g)^{1/2}d^{4}x},
\end{array}
\end{equation}
$$
\begin{array}{ccccc}
n\nu=-\displaystyle\frac{\partial{\mathcal L}}{\partial\varphi},&
\upsilon_{\mu}=\displaystyle\frac{(w\upsilon)_{,\mu}}{w}, &
\chi=\displaystyle\frac{\delta w}{w}, &
m^{2}=-\displaystyle\frac{w}{n}\displaystyle\frac{\partial^{2}{\mathcal
L}}{\partial\varphi^{2}},&
\Gamma=\displaystyle\frac{w}{n}\displaystyle\frac{\partial^{2}{\mathcal
L}}{\partial w\partial\varphi}.
\end{array}
$$
Here~$\overline{h}_{\mu\nu}=h_{\mu\nu}-\frac{1}{2}g_{\mu\nu}h$, is
the so-called tensor with inverse trace:
${\overline{h}=\overline{h}^{\sigma}{}_{\sigma}=-h=-h^{\sigma}{}_{\sigma}}$.
The operations of raising and lowering indices are performed using
the background metric~$g_{\mu\nu}$.

 The variable~$q$ is so remarkable,
because after linking all the potentials through
equations~(\ref{inverse-transformations1})~(\ref{inverse-transformations2})~(\ref{inverse-transformations3})
and substituting the~$q$-scalar~(\ref{qsc}) to the
expansion~(\ref{expansion}) we get a very simple perturbation
action~(\ref{lagrangian-q})~(totally divergent terms are
excluded), where~$q$ enters as a test massless-like field.

\section{Relation between $q$, $\zeta$ and $\Psi$}

Variables~$\zeta$ and $\Phi_{H}$ were introduced in
Bardeen~\cite{Bardeen 1} and Bardeen et al.~\cite{bardeen}. They
are expressed through~$q$ and $\Phi$ as follows:
\begin{equation}
\begin{array}{cc}
\zeta=\displaystyle\frac{2}{3}\frac{H^{-1}\dot{\Phi}+\Phi}{1+w_{B}}+\Phi=
\frac{1}{a\gamma}\frac{d}{dt}\left(\frac{a\Phi}{H}\right)=q, &
\Phi_{H}=\Phi,
\end{array}
\end{equation}
where (cf. (\ref{fridmann1}))
$$w_{B}=\frac{p}{\varepsilon},~~~\gamma=\frac{3}{2}(1+w_{B}).$$
Obviously $\zeta$ coincides with $q$ introduced in a general form
by equation (\ref{qsc}).

In the work~\cite{mukhanovreview} the dynamical equation
(equation~$(5.22)$ in~\cite{mukhanovreview}) was derived in
variables $\Psi$ and $\upsilon^{(gi)}$. In those notations the
variable~$q$~\cite{lukash} looks as follows:
\begin{equation}
\label{q-definition} q=\Psi+H\upsilon^{(gi)},
\end{equation}
where
\begin{equation}
\begin{array}{cc} \label{Psi-A} \Psi=A-\displaystyle\frac{(B'-C)a'}{a}, &
\upsilon^{(gi)}=\upsilon+a(B'-C).
\end{array}
\end{equation}

The identity of~(\ref{q-definition}) and (\ref{qsc}) is obvious if
we substitute~(\ref{Psi-A}) to~$(\ref{q-definition})$.

\section{Conclusion}

The theory of scalar cosmological perturbations can be constructed
in a quite general
form~(eqs.~(\ref{geometry})-(\ref{inverse-transformations3})).  In
order to get the Lagrangian for perturbations and the key
dynamical equation we needed the physical assumptions to
link~$\delta p_{c}$ and $\delta\varepsilon_{c}$. Moreover, the
assumption~(\ref{delta-p-c}) is equivalent to the assumption of a
matter field~$\varphi$ with a quite common Lagrangian~${\mathcal
{L}}(\varphi,w)$, where $\varphi$ is a 4-velocity potential of the
medium~(cf. (\ref{four-vel}) and (\ref{4-velocity})). In linear
theory of a single gravitating medium the two approaches coincide.

The variable~$\zeta$ introduced by Bardeen~et~al.~\cite{bardeen}
is equal to the variable~$q$ introduced by Lukash \cite{lukash},
and equation~$(5.22)$ found by
Mukhanov~et~al.~\cite{mukhanovreview} is equivalent
to~(\ref{qscalar}).

The author is grateful to V.~N.~Lukash, E.~V.~Mikheeva and
P.~B.~Ivanov for fruitful discussions. The work was supported by
UNK FIAN and Grant~{04-02-17444} of Russian Foundation for Basic
Research.

\end{document}